\documentclass{PoS}
\usepackage{txfonts}

\title{Ultra-peripheral-collision studies in the fixed-target mode with the proton and lead LHC beams}

\ShortTitle{UPC studies in the FT mode with $p$ and Pb LHC beams}

\author{\speaker{N.~Yamanaka}$^a$\thanks{Supported by JSPS Postdoctoral Fellowships for Research Abroad.}, C.~Hadjidakis$^a$, D.~Kiko{\l}a$^b$, J.P.~Lansberg$^a$,
        L.~Massacrier$^a$, M.G.~Echevarria$^c$, A.~Kusina$^d$, I.~Schienbein$^e$, J.~Seixas$^{f,g}$, 
        H.S.~Shao$^h$, A.~Signori$^i$, B.~Trzeciak$^j$, S.J.~Brodsky$^k$, G.~Cavoto$^l$,
        C.~Da~Silva$^m$, F.~Donato$^n$, E.G.~Ferreiro$^{o,p}$, I.~H\v{r}ivn\'{a}\v{c}ov\'{a}$^a$,
        A.~Klein$^m$, A.~Kurepin$^r$, C.~Lorc\'e$^s$, F.~Lyonnet$^t$, Y.~Makdisi$^u$,
        S.~Porteboeuf$^w$, C.~Quintans$^g$, A.~Rakotozafindrabe$^x$, P.~Robbe$^y$,
        W.~Scandale$^z$, N.~Topilskaya$^r$, A.~Uras$^{aa}$, J.~Wagner$^{ab}$, 
        Z.~Yang$^{ac}$, A.~Zelenski$^u$\\
        {\scriptsize
        \llap{$^a$}IPNO, CNRS-IN2P3, Univ. Paris-Sud, Universit\'e Paris-Saclay, 91406 Orsay Cedex, France\\
        \llap{$^b$}Faculty of Physics, Warsaw University of Technology, ul. Koszykowa 75, 00-662 Warsaw, Poland\\
        \llap{$^c$}Istituto Nazionale di Fisica Nucleare, Sezione di Pavia, via Bassi 6, 27100 Pavia, Italy\\
        \llap{$^d$}Institute of Nuclear Physics Polish Academy of Sciences, PL-31342 Krakow, Poland\\
        \llap{$^e$}Laboratoire de Physique Subatomique et de Cosmologie, Universit\'e Grenoble Alpes,
                   CNRS/IN2P3, 53 Avenue des Martyrs, F-38026 Grenoble, France\\
        \llap{$^f$}Dep. Fisica, Instituto Superior Tecnico, Av. Rovisco Pais 1, 1049-001 Lisboa, Portugal\\
        \llap{$^g$}LIP, Av. Prof. Gama Pinto, 2, 1649-003 Lisboa,Portugal\\
        \llap{$^h$}LPTHE, UMR 7589, Sorbonne University\'e et CNRS, 4 place Jussieu, 75252 Paris, France\\
        \llap{$^i$}Physics Division, Argonne National Laboratory, Lemont, IL 60439, USA\\
        \llap{$^j$}Institute for Subatomic Physics, Utrecht University, Utrecht, The Netherlands\\
        \llap{$^k$}SLAC National Accelerator Laboratory, Stanford University, Menlo Park, CA 94025, USA\\
        \llap{$^l$}``Sapienza" Universit\`a di Roma, Dipartimento di Fisica \&
                   INFN, Sez. di Roma, P.le A. Moro 2, 00185 Roma, Italy\\
        \llap{$^m$}P-25, Los Alamos National Laboratory, Los Alamos, NM 87545, USA\\
        \llap{$^n$}Turin University, Department of Physics, and INFN, Sezione of Turin, Turin, Italy\\
        \llap{$^o$}Dept. de F{\'\i}sica de Part{\'\i}culas \& IGFAE, Universidade de Santiago de Compostela,
                   15782 Santiago de Compostela, Spain\\
        \llap{$^p$}Laboratoire Leprince-Ringuet, Ecole polytechnique, CNRS/IN2P3, Universit\'e Paris-Saclay, Palaiseau, France\\
        \llap{$^r$}Institute for Nuclear Research, Moscow, Russia\\
        \llap{$^s$}CPHT, \'Ecole Polytechnique, CNRS,  91128 Palaiseau, France\\
        \llap{$^t$}Southern Methodist University, Dallas, TX 75275, USA\\
        \llap{$^u$}Brookhaven National Laboratory, Collider Accelerator Department\\
        \llap{$^w$}Universit\'e Clermont Auvergne, CNRS/IN2P3, LPC, F-63000 Clermont-Ferrand, France\\
        \llap{$^x$}IRFU/DPhN, CEA Saclay, 91191 Gif-sur-Yvette Cedex, France\\
        \llap{$^y$}LAL, Universit\'e Paris-Sud, CNRS/IN2P3, Orsay, France\\
        \llap{$^z$}CERN, European Organization for Nuclear Research, 1211 Geneva 23, Switzerland\\
        \llap{$^{aa}$}IPNL, Universit\'e Claude Bernard Lyon-I and CNRS-IN2P3, Villeurbanne, France\\
        \llap{$^{ab}$}National Centre for Nuclear Research (NCBJ), Ho\.{z}a 69, 00-681, Warsaw, Poland\\
        \llap{$^{ac}$}Center for High Energy Physics, Department of Engineering Physics, Tsinghua University, Beijing, China\\}
        }

\abstract{
We address the physics case related to the studies of ultra-peripheral $p$H, $p$Pb, PbH,  and PbPb collisions in the fixed-target mode at the LHC. 
In particular, we discuss how one can measure the gluon generalized parton distribution $E_g(x,\xi,t)$ in exclusive $J/\psi$ photoproduction with a transversely polarized hydrogen target.
}

\FullConference{International Conference on Hard and Electromagnetic Probes of High-Energy Nuclear Collisions\\
		30 September - 5 October 2018\\
		Aix-Les-Bains, Savoie, France}

\begin{document}

\section{Introduction}

Ultra-peripheral collisions (UPC) of hadrons or nuclei, whereby these interact via at least a photon exchange, allows one to use hadron colliders in order to probe the properties of quantum chromodynamics (QCD) in similar ways than at lepton-hadron colliders~\cite{Baltz:2007kq}. 
The almost on-shell photons coherently emitted by these fast moving relativistic charges allow one, among other things, to study a variety of photoproduction reactions. 
UPC are now routinely studied at the LHC and RHIC colliders and they will progressively start to be used to extract information on the hadron structure. 

In this proceedings contribution, we focus on the case of the hadron LHC beams used in the fixed-target (FT) mode~\cite{Brodsky:2012vg,Hadjidakis:2018ifr} and how UPCs can allow one to study exclusive photo-production of quarkonia $\gamma p \to {\cal Q} p$~\cite{Lansberg:2018fsy,Massacrier:2017lib}, which are known to be a good probe of the tri-dimensional gluon content of the proton. 
In such studies, the transverse polarization of the proton target would also provide a direct access to the gluon orbital angular momentum which has never been measured so far.

\section{Kinematics of the ultra-peripheral collisions}

UPCs, by definition, happen when the impact parameter $b$ of the colliding system is larger than the sum of the radii.
The typical maximal energy of the photon (in the rest frame of the emitter) is given by 
$
E_\gamma^{\rm max, rest} 
\simeq \frac{1}{R_X + R_Y}$, where $R_X$ and $R_Y$ are the radii of two colliding nuclei. For collisions involving lead ions, one has $E_\gamma^{\rm max, rest}  \simeq 30\,$ MeV. This energy has then to be boosted in the appropriate frame. 

\begin{table}[!htb]
\begin{center}
\caption{
The key figures of the UPC in the FT mode at the LHC \cite{Lansberg:2015kha}.
$E_{\rm beam}$ is the beam energy per nucleon.
}
\begin{tabular}{lccc}
\hline
System & $E_{\rm beam}$ [GeV] & $\sqrt{s_{\gamma p}^{\rm max}}$ [GeV] & $\sqrt{s_{\gamma \gamma}^{\rm max}}$ [GeV] \\
\hline
$p$H & 7000 & 44 & 17\\
$p$Pb & 7000 & 19 & 3.2 \\
PbPb & 2760 & 9 & 1.0 \\
PbH & 2760 & 12 & 3.2 \\
\hline
\end{tabular}
\label{table:key_figures}
\end{center}
\end{table}

In the FT case, for one-photon exchanges, the Lorentz factor, $\gamma_L$, is simply the ratio of the emitter energy $E$ in the rest frame of the receiver, divided by its mass. As such, the maximal center-of-mass (CM) energy of the photon-hadron collision 
\begin{equation}
\sqrt{s_{\gamma p}^{\rm max}}
=
\sqrt{2 \times \gamma_L \times  E_\gamma^{\rm max,rest } \times m_N}.
\end{equation}
We note that the photon can in principle be radiated by both the projectile and the target. For hard reactions, the fact
that the target is not ionized does not matter.
We give in Table \ref{table:key_figures} the key figures of the UPC in FT experiments at LHC.
We see that $\sqrt{s_{\gamma p}^{\rm max}}$ is high enough to create charmonia ($\ge 4$ GeV). In fact, such energies are
high enough to perform the first UPC studies in the FT mode ever.

\section{Gluon orbital angular momentum and generalized parton distributions}

It is now an evidence that the proton spin is not saturated by the quark and gluon spin, but the remaining piece, which should be filled by the parton Orbital Angular Momentum (OAM), is still obscure, in particular that of the gluon.
The gluon OAM is indirectly related to the gluon Sivers function, which was recently measured by COMPASS through the semi-inclusive deep inelastic scattering \cite{Adolph:2017pgv}.
The result is showing a nonzero Sivers asymmetry around the gluon momentum fraction $x_g \simeq 0.1$, and it is providing an important hint for nonzero gluon OAM. The gluon Sivers effect may also be probed via single transverse spin asymmetry (STSA) of single-inclusive quarkonia, $A_N^{\rm incl. onium}$. We refer to \cite{Hadjidakis:2018ifr,Kikola:2017hnp} for performance studies in the LHC FT modes and for a contextual discussion. 

As an alternative observable to access the gluon OAM, we have the Generalised Parton Distribution (GPD). 
From Ji's sum rule, 
the gluon OAM is obtained by integrating the GPDs \cite{Ji:1996ek}:
\begin{equation}
{\cal L}^g
=
\int x_g dx_g \,[H^g (x_g,0,0) +E^g (x_g,0,0)]
- \int dx_g \tilde{H}^g (x_g,0,0)
,
\end{equation}
where $H^g$, $E^g$, and $\tilde{H}^g$ are the twist-2 GPDs.

\begin{figure}[hbt]
\begin{center}
\includegraphics[clip,width=.3\columnwidth]{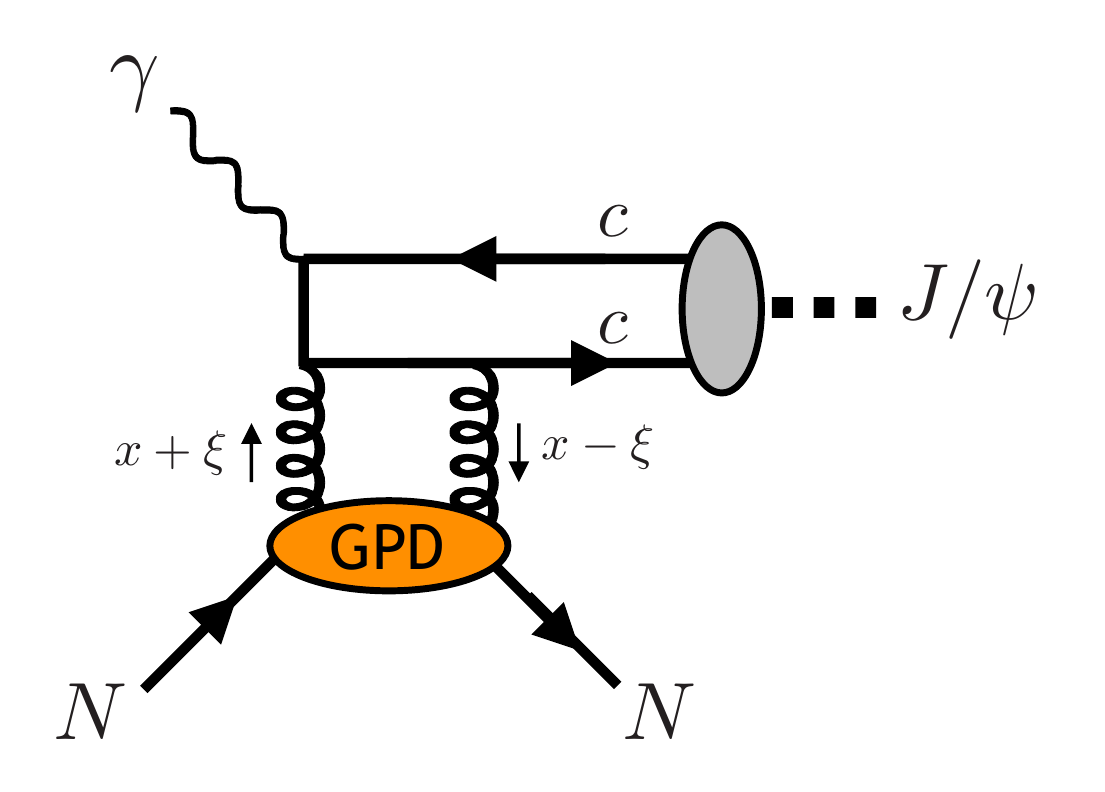}
\vspace{-1em}
\caption{
Diagrammatic representation of the exclusive $J/\psi$ production through the nucleon GPD.
}
\label{fig:Jpsi_exclusive_GPD}
\end{center}
\end{figure}

The gluonic GPDs $H^g$ and $E^g$ may be extracted from the exclusive $J/\psi$ photo-production off a proton.
Its leading order contribution to the amplitude is given by
\begin{equation}
{\cal M}
\propto
\left[ {\cal H}^g \bar u (p') n\hspace{-0.5em}/_- u(p)
+{\cal E}^g \bar u (p') \frac{i \sigma^{n_{-} \Delta}  }{2m_N} u(p) \right]
,
\end{equation}
where ${\cal H}^g (\xi , t ) \equiv \int_{-1}^1 dx \, \frac{\alpha_s \xi H^g (x, \xi ,t) }{(x-\xi +i\epsilon)(x+\xi -i\epsilon)} $ and ${\cal E}^g (\xi , t ) \equiv \int_{-1}^1 dx \, \frac{\alpha_s \xi E^g (x, \xi ,t) }{(x-\xi +i\epsilon)(x+\xi -i\epsilon)}$ with $x$ and $\xi$ the momentum fraction of the gluon in the proton and that exchanged through the transition between initial and final states to define the GPD (see Fig. \ref{fig:Jpsi_exclusive_GPD} for the definition).
We also used $n_-$ and $t\equiv \Delta^2$ to denote the light-cone momentum (chosen so that the initial nucleon momentum is decomposed as $p=(1+\xi) \sqrt{s_{\gamma p}} n_+ + \frac{m_N^2}{2(1+\xi ) \sqrt{s_{\gamma p}}} n_-$) and the squared exchanged momentum, respectively.
We note that to access ${\cal E}^g$, we need to transversely polarize the proton target.
Indeed, the STSA for exclusive $J/\psi$ production reads~\cite{Koempel:2011rc,Lansberg:2018fsy}:
\begin{equation}
A_N^{\gamma p^{\uparrow} \to J/\psi\, p}
\propto
(1+\xi)
\, {\rm Im} \left( {\cal E}^{g*}{\cal H}^{g} \right)
.
\end{equation}

\section{Projections for $A_N^{\gamma p^{\uparrow} \to J/\psi\, p}$ measurements in the LHC FT mode}

\begin{figure}[hbt]
\begin{center}
\includegraphics[width=.49\columnwidth]{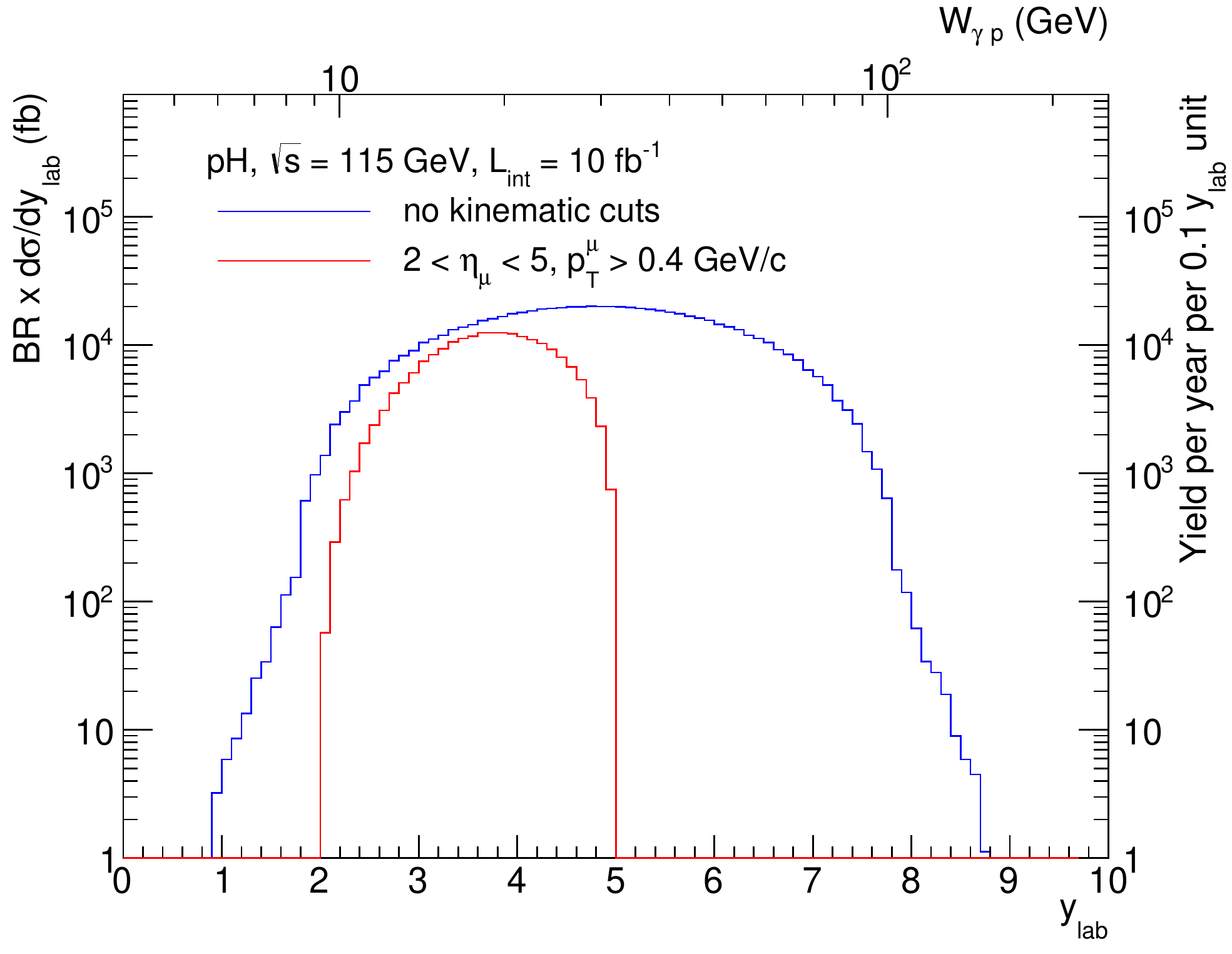}
\includegraphics[width=.49\columnwidth]{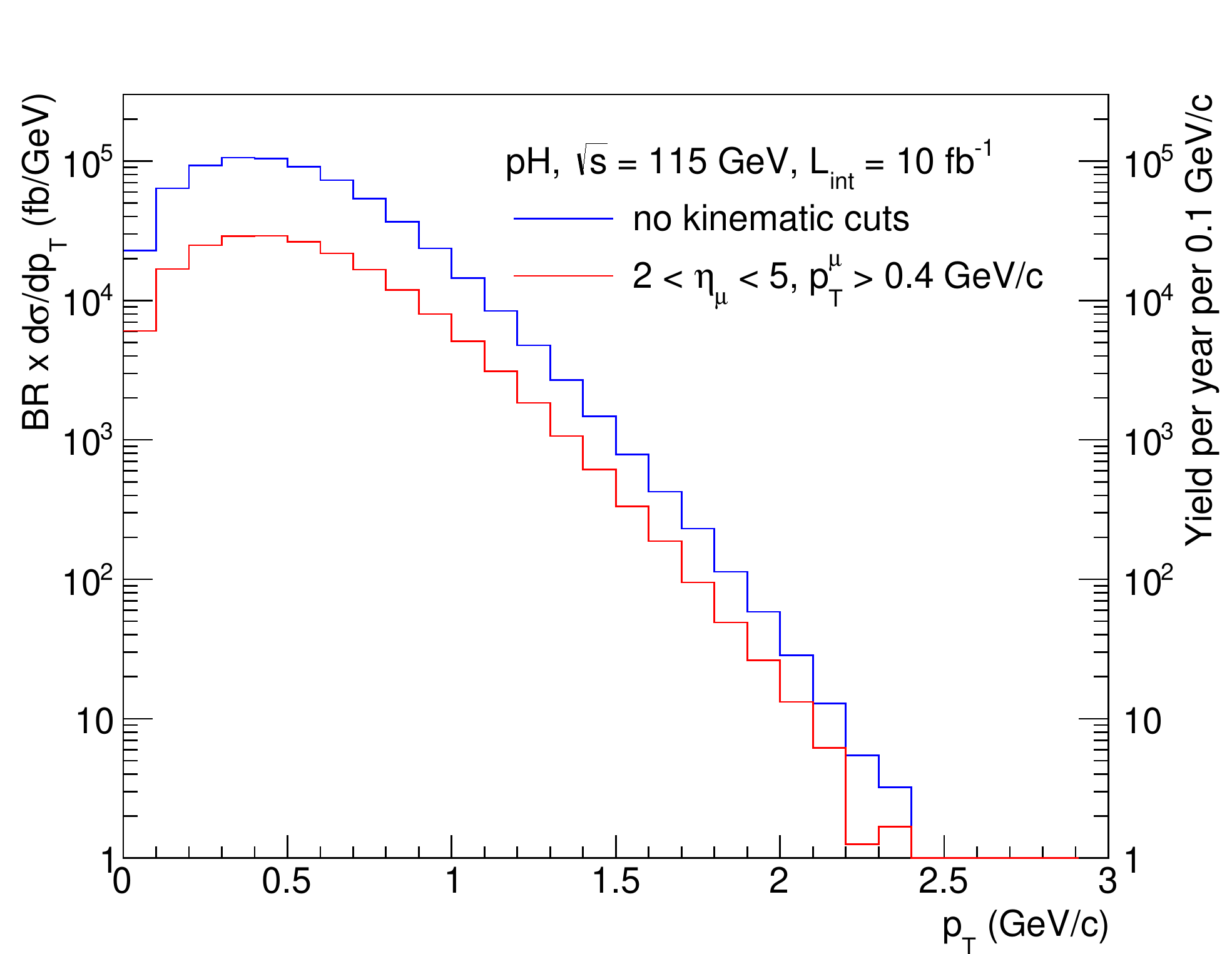}
\includegraphics[width=.49\columnwidth]{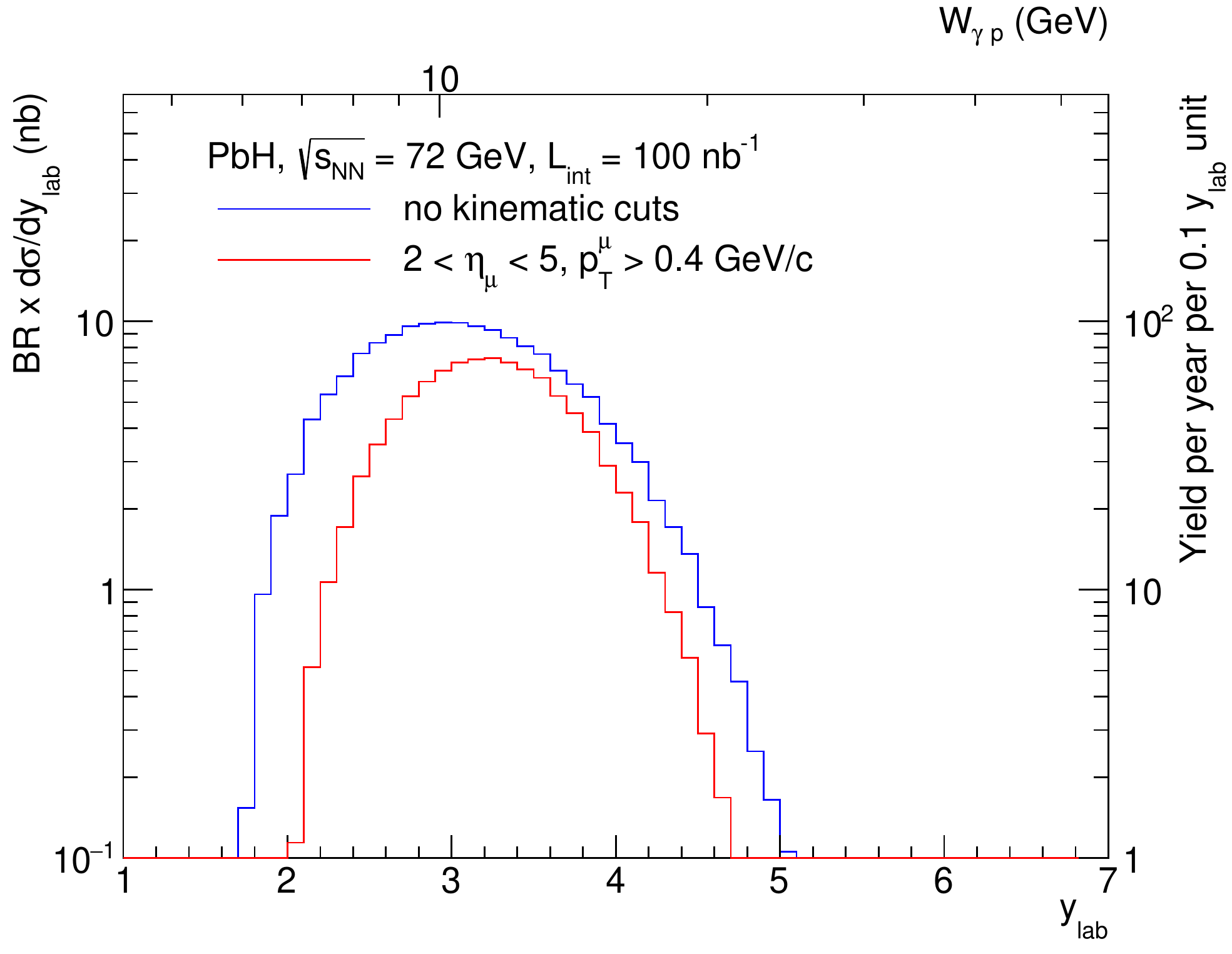}
\includegraphics[width=.49\columnwidth]{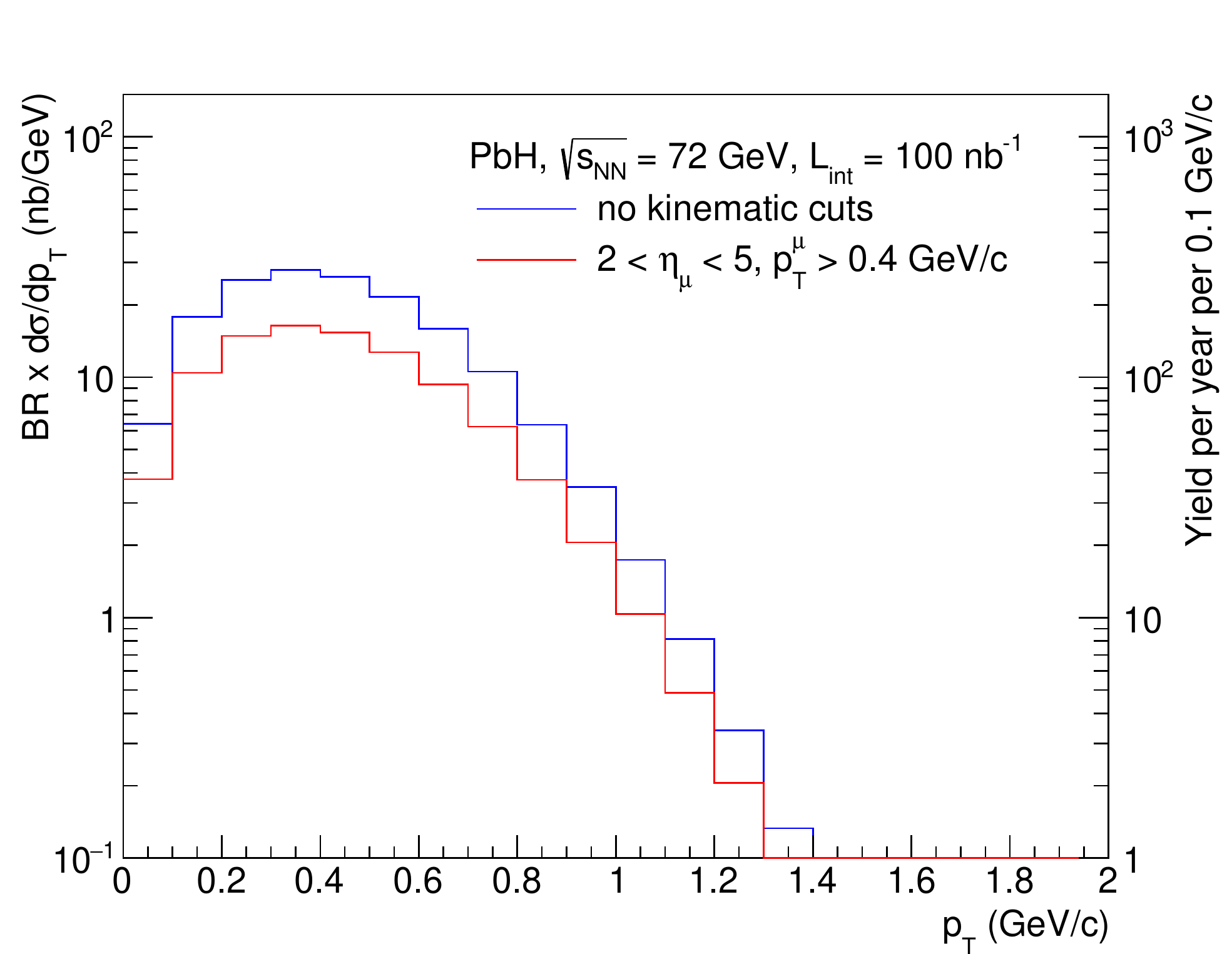}
\caption{
Rapidity ($y$, left panels) and transverse momentum ($p_T$, right panels) differential cross section of the exclusive $J/\psi$ production via UPCs for the case of $p$H (upper panels) and PbH (lower panels). From~\cite{Lansberg:2018fsy}.
}
\label{fig:AFTER_pH_PbH}
\end{center}
\end{figure}

We report now on simulations~\cite{Lansberg:2018fsy} of the exclusive $J/\psi$ production off proton via UPCs in the LHC FT mode.
The simulation was performed using {\small \sc STARLIGHT}~\cite{Klein:2016yzr} with the kinematical cuts of LHCb (muon pseudorapidity $2 <\eta_\mu < 5$ and muon transverse momentum $p_T^\mu >0.4$ GeV) being applied.
The cross section as well as the number of events expected from one-year run at LHCb are displayed in Table \ref{table:jpsi_events}.
We also show in Fig. \ref{fig:AFTER_pH_PbH} the differential cross section for the cases of $p$H and PbH collisions.

\begin{table}[htb]
\begin{center}
\caption{
Result of the simulations of $\gamma p^{\uparrow} \to J/\psi\, p$ via UPC in the FT mode with the LHCb setup.
The luminosity assumed for the $p$H (PbH) case is ${\cal L}=1.0\times 10^4$ pb$^{-1}$yr$^{-1}$ ($0.12$ pb$^{-1}$yr$^{-1}$).
}
\begin{tabular}{lcc}
\hline
& $p$H & PbH\\
\hline
Photon-emitter & $p$ & Pb \\
$\sigma_{J/\psi}^{\rm tot}$ (pb) & $1.2 \times 10^3$ & $2.8 \times 10^5$ \\
$\sigma_{J/\psi \to l^+ l^-}$ (pb) & 70 & $1.7 \times 10^4$ \\
$\sigma_{J/\psi \to l^+ l^-}$ (with LHCb cuts) (pb) & 21 & $9.8 \times 10^3$ \\
Number of events (1 year) & $2\times 10^5$ & $1 \times 10^3$ \\
\hline
\end{tabular}
\label{table:jpsi_events}
\end{center}
\end{table}

\begin{figure}[hbt!]
\begin{center}
\includegraphics[width=.49\columnwidth]{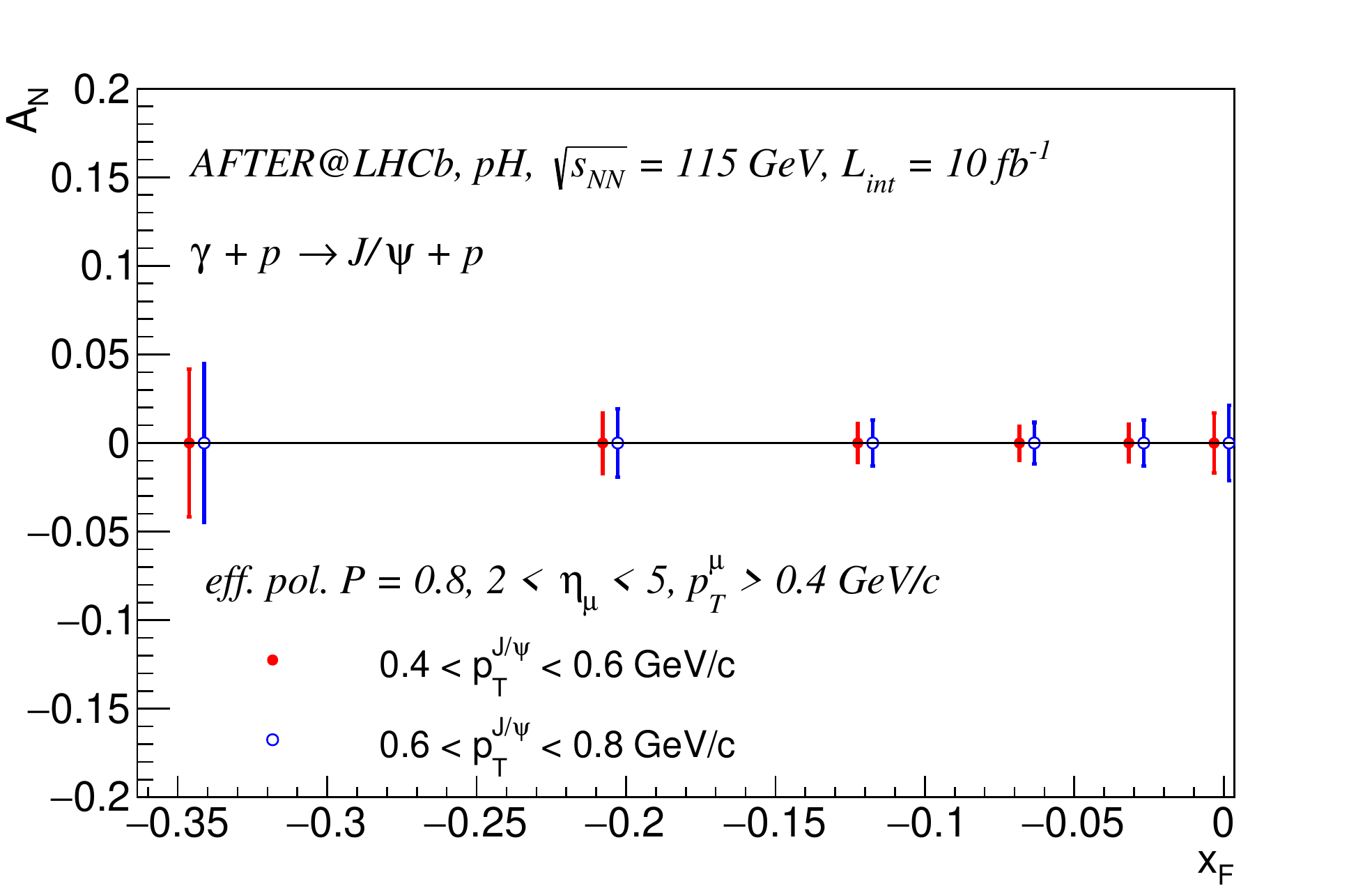}
\includegraphics[width=.49\columnwidth]{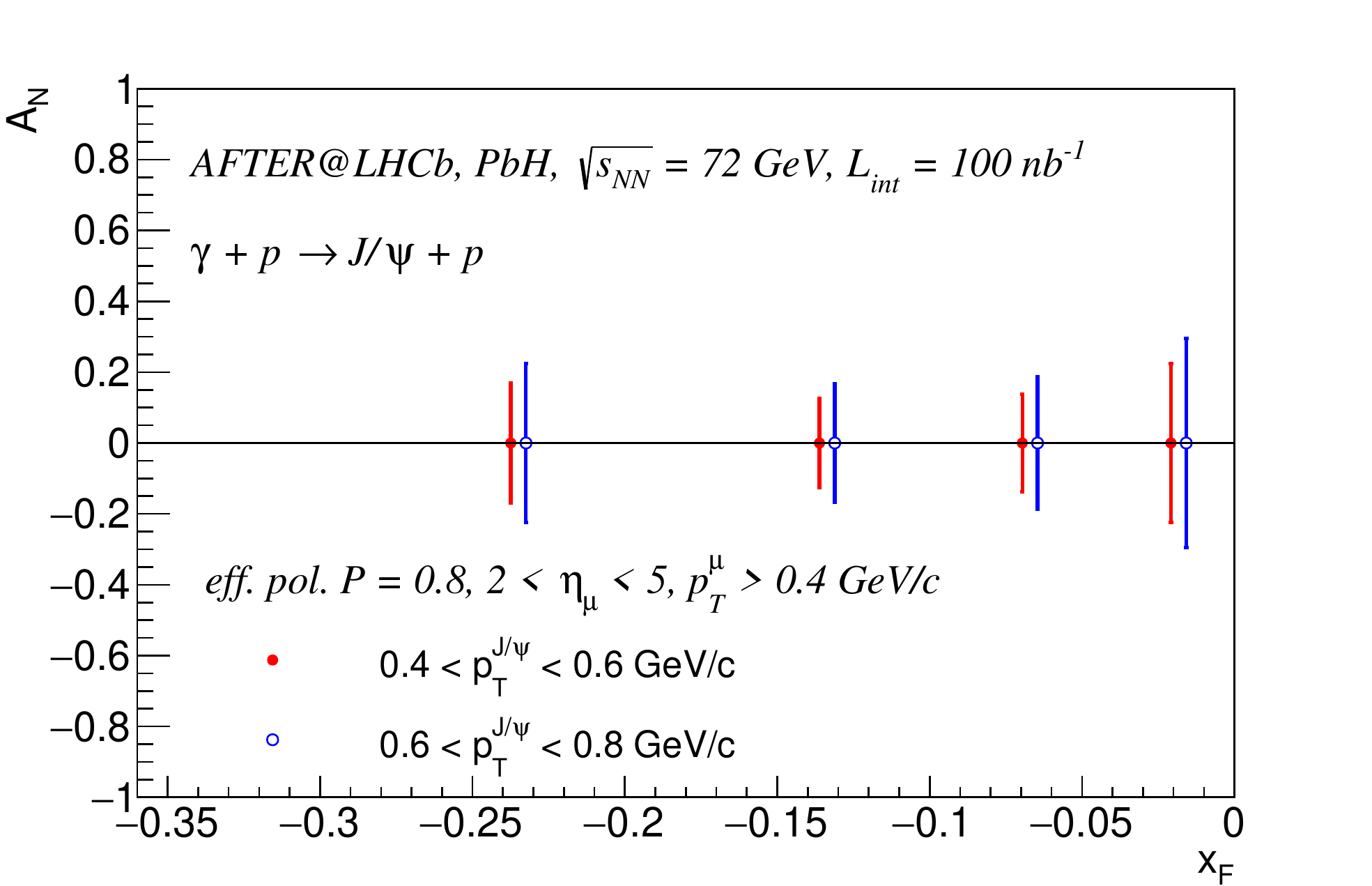}
\caption{
Statistical uncertainty projections for the STSA in exclusive $J/\psi$ production via UPCs for the case of $p$H (left panel) and PbH (right panel). From~\cite{Lansberg:2018fsy}.
}
\label{fig:assym-mod}
\end{center}
\end{figure}

By assuming a target polarization of 80\%, compatible with the H-jet system from RHIC or HERMES-like storage cell (see~\cite{Hadjidakis:2018ifr} for a complete discussion of the possible implementations), the sensitivity of the FT experiment on the STSA was projected, as shown in Fig. \ref{fig:assym-mod}.
We note that the proton is not detected in the process $\gamma p \to J/\psi  p$.
We also considered the detector as fully efficient for the dimuon detection after the kinematic cuts in the sensitivity studies for $A_N$.
The projection indicates that $A_N$ for the pH case can be determined within few percents, and 10\% for that of the PbH collisions.
We note that the latter is better from the point-of-view of the UPC, since the source of the emitted photons is nearly always the lead ion.
It would also be interesting to study the case of the deuteron target, which will allow us to access to the deuteron GPD~\cite{Berger:2001zb,Cano:2003ju,Cosyn:2018rdm}.

\section{Conclusion and outlook}

In this proceedings contribution, we reported on the potentiality of the LHC FT mode to perform the first study of UPCs in the FT mode. 
It was shown that it can be used to probe the gluon GPDs via quarkonium exclusive photoproduction.
Using a transversely polarized target, one can access $E^g$ and thus the gluon orbital angular momentum.
With 10 fb$^{-1}$($p$H), it is possible to measure $A_N^{J/\psi}$ with a few \% of accuracy.

The experimental investigation of the UPC may also provide other interesting informations such as the quark GPDs, which is accessible through exclusive dilepton photoproductions.
The interference between the Bethe-Heitler and Timelike Compton scatterings gives indeed access to the quark GPDs via the azimuthal asymmetry, and
the study for the cases of $p$Pb, PbH, and $p$H systems shows that
the interference can be as large as 10\% of the total cross section for the energy in the FT mode of LHC experiment \cite{Lansberg:2015kha}.

Another interesting possibility is to measure the light-by-light scattering, which was recently observed by the ATLAS and CMS Collaborations \cite{Aaboud:2017bwk,Sirunyan:2018fhl}.
In addition, the $\gamma \gamma$ CM energy in the FT mode exceeds the charmonium production threshold for $p$H, $p$Pb, and PbH (see Table \ref{table:key_figures}), so that we may also measure the $\eta_c$ photoproduction \cite{Goncalves:2015hra,Klein:2018ypk} or the production axions with a mass in the MeV-GeV region \cite{Knapen:2016moh}.
The former will permit us to quantify the photon flux of the projectile (not possible for the PbPb case due to the lack of CM energy).
The latter aims at scanning the mass range of axion-like particles in which the constraint on the axion-photon coupling is still loose \cite{Jaeckel:2015jla}.
The feasibility studies for these processes using the FT experiments remain to be done in the future.

\end{document}